\documentclass[12pt,oneside]{article}
\usepackage{amsmath,amssymb,graphics}
\usepackage[left=3.2cm,top=3.6cm,bottom=3.6cm,right=3.2cm,head=0cm,foot=0.7cm]{geometry}
\newcommand{\field}[1]{\mathbb{#1}}

\usepackage[font=small]{caption}

\usepackage[usenames,dvipsnames]{color}

\title{Can Somebody Please Say What Gibbsian Statistical Mechanics Says?}

\author{Roman Frigg\footnote{London School of Economics, romanfrigg.org, r.p.frigg@lse.ac.uk.}  $\,$ and Charlotte Werndl\footnote{University of Salzburg and London School of Economics, charlotte.werndl@sbg.ac.at, charlottewerndl.net.}\vspace{0.5cm}}

\date{{\small Forthcoming in \emph{The British Journal for the Philosophy of Science}}}

\begin{document}

\maketitle

\begin{abstract}

\noindent Gibbsian statistical mechanics (GSM) is the most widely used version of statistical mechanics among working physicists. Yet a closer look at GSM reveals that it is unclear what the theory actually says and how it bears on experimental practice. The root cause of the difficulties is the status of the Averaging Principle, the proposition that what we observe in an experiment is the ensemble average of a phase function. We review different stances toward this principle, and eventually present a coherent interpretation of GSM that provides an account of the status and scope of the principle. 

\end{abstract}

\newpage
\tableofcontents
\newpage


\section{Introduction} \label{Intro}

Gibbsian statistical mechanics (GSM) is a  powerful theory because it offers a general prescription of how to calculate an array of equilibrium properties, and its methods are applied across a large range of different systems. For these reasons GSM is widely used and considered by many to be \emph{the} formalism of statistical mechanics.\footnote{The first chapter of Landau and Lifshitz's ([1980]) classical introduction is entitled `the fundamental principles of statistical physics' and is dedicated entirely to a discussion of the Gibbs formalism; Isihara ([1971]) introduces the Gibbs formalism in a chapter called `principles of statistical mechanics'. These are no exceptions.}  Yet a closer look at GSM reveals that it is unclear what the theory says and how it bears on experimental practice. The issue is that GSM is an ensemble theory and as such it makes statements about the behaviour of ensembles rather than about individual physical systems. A common way of bridging the gap between ensembles and systems is to adopt an averaging principle: what we observe in an experiment on a system is the ensemble average of a phase function representing a relevant physical quantity. Although this principle is commonly used in applications, the status and scope of the  principle are unclear and different authors make conflicting pronouncements. But until the status and scope of the averaging principle are settled, or some alternative method of bridging the gap between ensembles and individual systems has been found, it is unclear what the content of GSM is. Hence our plea: can somebody please say what GSM says?\\

\noindent We first introduce GSM, lay bare the problem of connecting the formalism of GSM to the behaviour of individual physical systems, and document the lack of a consensus on the matter in the physics literature (Section \ref{Ensembles}). We then point out that remaining silent about the problem does not make the problem go away (Section \ref{Quietism}), and that the traditional answer in terms of time averages is not viable for a number of reasons (Section \ref{TAA}). We proceed to introduce a position we call `bare probabilism', the view that the probability measures in GSM give universally true probabilities for events to occur and that nothing else should be read into GSM. We discuss two versions of this view, one based on ensembles and one based on the notion that the theory's probability measure $\rho$ is the macro-state of a system, and point out that both of them leave essential questions unanswered (Section \ref{Probabilism1}). The answer to these questions comes from the study of fluctuations. These can be integrated into probabilism in two different ways, leading to two equally acceptable positions we call `qualified probabilism' and `fluctuation probabilism' (Section \ref{Fluctuations}). Probabilism endorses \emph{$\rho$-universalism}, the view that $\rho$  gives the correct probabilities for all events at all times. We point out that this  assumption is not generally true and that probabilism only works if what we call the `masking condition' or the `$f$-independence condition' hold. Under these conditions, and if fluctuations turn out to be thermodynamic, the averaging principle is justified (Section \ref{Universalism}). We finally draw some conclusions and point out that there are justifications of the averaging principle that do not rely on probabilism (Section \ref{Conclusion}). \\

\noindent Two caveats are in place. First, throughout the paper we discuss GSM in the setting of classical mechanics, and we assume that the dynamics is deterministic. We believe that many of the issues we discuss will recur when GSM is applied to classical stochastic systems or to quantum systems,\footnote{Classical stochastic systems are discussed in Werndl and Frigg ([2017b]).} but space constraints prevent us from discussing these systems here. Second, even though there are attempts to generalise GSM to non-equilibrium situations, GSM is first and foremost an equilibrium theory. We restrict attention to the equilibrium case and set non-equilibrium considerations aside.


\section{Systems and Ensembles}\label{Ensembles}

A \emph{system} $S$ is a part of the physical world: a gas in a container, a crystal on the laboratory table, a liquid in jar. From a mathematical point of view a system has the structure of a so-called \emph{measure-preserving dynamical system}. In the current context the difference between a physical system and its mathematical representation is inessential, and we use the term `system' to refer to either the physical object or its mathematical representation.\footnote{For a discussion of what it takes for mathematical object to represent a physical system see Frigg and Nguyen ([2018]).}  A system, then, is a quadruple $(X, \Sigma_{X}, \phi_{t}, \mu)$. $X$ is the system's \emph{state space}, which contains all states the system's \emph{micro-constituents} could assume. For this reason the states in $X$ are referred to as \emph{micro-states}. In the case of a gas consisting of $n$ molecules, $X$ has $6n$ dimensions: three dimensions for the position of each particle and three dimensions for the corresponding momenta. $\Sigma_{X}$ is a $\sigma$-algebra of subsets of $X$, and $\mu$ is a measure on $\Sigma_{X}$. The \emph{evolution function} $\phi_{t}$ determines how the system's micro-state changes over time. If at a certain time $t_{0}$ the system is in micro-state $x_{0}$, then it will be in state $\phi_{t}(x_{0})$ at a later time $t$. The path that $\phi_{t}(x_{0})$ traces through $X$ as time evolves is the \emph{trajectory through $x_{0}$}, and $x_{0}$ is the \emph{initial condition}. Throughout we assume that $\mu_{X}$ is invariant under $\phi_{t}$, meaning that the subsets of $X$ can change their shape but not their measure as time evolves.\footnote{A more extensive description of the elements of a system can be found in Werndl and Frigg ([2015]). For a general introduction to dynamical systems see Katok and Hasselblatt ([1993]).}\\ 

\noindent At the macro level the system is characterised by a set of \emph{macro-variables} such as volume, pressure, and magnetisation. From a mathematical point of view macro-variables are functions $f: X \rightarrow \field{R}$; i.e. functions that associate a real number with each point in $X$. If, for instance, $f$ is the magnetisation of the system and the system is in micro-state $x$, then $f(x)$ is the magnetisation of the system when it is in micro-state $x$.\\ 

\noindent Systems as defined so far do \emph{not} come equipped with a notion of equilibrium; nor is any such notion tacitly implied either by the definition of a system or by the definition of a macro-variable.\\

\noindent The key concept of GSM is that of an \emph{ensemble}. Informally an ensemble is often described as an infinite collection of systems of the same kind (and with the same time evolution) that differ in their state.\footnote{This is Gibbs' original characterisation ([1981], p. v, p. 5). See also Agarwal and Eisner ([1988], p. 5), Hill ([1986], p. 3, [1987], p. 4), Kittel ([2004], pp. 7--8), and Schr\"odinger ([1989], p. 3).} It is important that an ensemble is a collection of copies of the \emph{entire} system $S$ (as characterised above) and not a collection of molecules. In fact, an ensemble is best thought of as a collection of `mental copies of the one system under consideration' (Schr\"odinger [1989], p. 3). Hence the members of an ensemble do not interact with each other; an ensemble is not a physical object; and ensembles have no spatiotemporal existence.\\ 

\noindent A natural question to ask about a collection of systems that differ in their instantaneous state is how these states are distributed. The distribution of states is formalised with the aid of a probability measure $\rho$ on $X$, which encodes, intuitively speaking, what proportion of systems in the collection are in states that are located in a certain subset of $X$. With $\rho$ at hand, one would now want to ascend from the intuitive to the precise and offer a formal definition of an ensemble. But at this point an ambiguity becomes apparent. The issue is that talk of a `collection of systems' allows for two readings.\\ 

\noindent On a narrow reading, an ensemble is a collection of systems that are \emph{distributed in a particular way}. This can be formalised by associating the ensemble with a \emph{particular} probability measure: the ensemble simply \emph{is} the measure $\rho$. This squares with locutions like `the microcanonical ensemble' or `the canonical ensemble', which seem to associate ensembles with a particular distribution (more about these distributions below). However, this reading has an important limitation. In the physics literature one also often encounters talk of the \emph{state of an ensemble}, and of the state of an ensemble changing.\footnote{See, for instance, Tolman ([1979], p. 46).} But talk of the state of an ensemble is pointless if the ensemble is \emph{defined} by $\rho$: if the distribution changes, one ensemble goes out of existence and a new ensemble is created. On the narrow reading, a change in $\rho$ is not a change of an ensemble's state; it is an act of annihilation and recreation.\\ 

\noindent A broad reading accommodates the notion of a change of state. It interprets an ensemble as a `bare' collection of systems: a collection of systems that has \emph{no particular distribution}. The systems in the ensemble can then be distributed in different ways. These ways can be described by different measures $\rho$, which are the ensemble's states. This suggest a definition of an ensemble as a pair $(X, P)$, where $P$ is the class of all probability measures over $(X, \Sigma_{X})$. $P$ is the ensemble's state space: each measure in $P$ specifies a state of the ensemble; if the measure changes, the ensemble's state changes. The time evolution of the ensemble is described by changes in $\rho$. If changes in $\rho$ happen solely due the unperturbed evolution of the systems in the ensemble, then the ensemble distribution at time $t$ is given by $\rho_{t}(x) = \rho_{0}(\phi_{-t}(x))$, where $\rho_{0}$ is the distribution at the initial time $t_{0}$.\\

\noindent Nothing in what follows depends on which of the two definitions of an ensemble one adopts, and our points can be made with either of the definitions. However, our discussion is more easily framed in the broader reading, and hence we adopt that reading from now on.\\

\noindent The \emph{ensemble average} of a macro-variable  $f$ for an ensemble in state $\rho$ is 

\begin{equation}\label{EA}
\langle f \rangle  = \int_{X}f(x)\rho(x, t) dx.
\end{equation}

\noindent In his discussion of ensembles Gibbs introduces what he calls the `condition of statistical equilibrium' ([1981], p. 8). An ensemble is in \emph{statistical equilibrium} iff (if and only if) $\rho$ is stationary.\footnote{This view is adopted widely in the extant physics literature as well as in philosophical discussions about GSM; see, for instance, Hill ([1987], p. 8), Myrvold ([2016], pp. 588--9), and Tolman ([1979], p. 63). A diverging account was put forward by van Lith ([1999]) who suggested replacing stationarity with the requirement that the distribution be such that the phase average of a few selected phase functions be constant. Our concerns are orthogonal to van Lith's and none of the issues that we discuss below change if van Lith's definition is adopted.} A distribution is stationary iff it does not change over time, meaning that it is invariant under the dynamics: $\rho_{0}(x) = \rho_{t}(x)$ for all $t$. For obvious reasons $\langle f \rangle$ is constant if the distribution is stationary.\\

\noindent There will usually be a large number of stationary distributions for a certain dynamics, and so the question arises which of these distributions should be chosen to characterise a given situation. Gibbs discusses this issue at length and proposes the so-called microcanonical distribution if the system is completely isolated from its environment, and the so-called canonical distribution if the system is in contact with a heat bath of a certain temperature.\footnote{These distributions are discussed in any textbook on statistical mechanics. See for instance, Lavis ([2015]).} The justification of these distributions as the correct distributions for certain situations can proceed along different lines, and a number of suggestions have been made.\footnote{For surveys see Frigg and Werndl ([forthcoming]) and Myrvold ([2016]).} We pass over this matter here because nothing in what follows depends on how the choice of a particular distribution is justified. The crucial point to bear in mind is that statistical equilibrium pertains to an ensemble and hence provides a notion of an \emph{ensemble-equilibrium}.\\

\noindent The notion of an ensemble-equilibrium contrasts with that of a system-equilibrium. As noted above, a system is part of the physical world. When is such a part in equilibrium? Thermodynamics (TD) offers a clear answer to this question: a system is in \emph{thermodynamic equilibrium} `when none of its thermodynamic properties are changing with time' (Reiss [1996], p. 3).\footnote{Thermodynamic equilibrium is also discussed in Buchdahl ([1975], p. 2), Fermi ([2000], p. 4), Guggenheim ([1967], p. 7), Gyftopoulos ([2005], p. 55), Honig ([1999], p. 3), Kubo ([1968], p. 2), Pippard ([1966], pp. 5--6), and van Ness ([1969], p. 74). Being in thermodynamic equilibrium is an intrinsic property of the system, which offers a notion of  `internal equilibrium' (Guggenheim [1967], p. 7). It contrasts with `mutual equilibrium' (\emph{ibid.}, 8), which is the relational property of \emph{being in equilibrium with each other}  that two systems eventually reach after being put into thermal contact with each other. Mutual equilibrium is often referred to as `thermal equilibrium'. It is the notion that figures in the zeroth law of thermodynamics, which effectively says that thermal equilibrium is an equivalence relation. In the current context, thermodynamic  rather than thermal equilibrium is the relevant concept.} In other words, the idea is that we look at a full set of thermodynamic variables and track their values over time; when the values do not change any more, then the system has reached equilibrium. The values that the system settles on when values do not change any more are the \emph{equilibrium values}. \\

\noindent We are now faced with two different objects of study and two different notions of equilibrium. On the one hand there are  systems, which can be in thermodynamic equilibrium. On the other hand there are ensembles, which can be in statistical equilibrium. How do these points of view relate to one another? This question is of paramount importance because statistical mechanics requires both. A physical system is what one has experimental access to and what one ultimately aims to study. At the same time it is the ensemble approach that provides the mathematical apparatus to carry out calculations and generate predictions. But what do calculations performed on an \emph{ensemble} tell us about the properties of a \emph{system}? Or for those who prefer to think about the problem in terms of experiments: what do ensemble calculations tell us about experimental results? \\ 

\noindent A common way to bridge the gap between ensembles and systems is to adopt what we call the \emph{Averaging Principle} (AP): when measuring the property $f$ on a system in system-equilibrium, the observed equilibrium value of the property is the ensemble average $\langle f \rangle$ of an ensemble in ensemble-equilibrium.\footnote{AP only makes the conditional claim that if a physical quantity is associated with a phase function $f$, then the outcome of a measurement of that quantity is $\langle f \rangle$. It remains silent about temperature and entropy, for which there are no phase functions.} About two thirds of the around thirty textbooks on statistical mechanics we consulted when researching this paper offer explicit statements of AP, and those that do not explicitly state AP appeal to it tacitly by basing their considerations on averages.\footnote{Explicit statements are given by Agarwal and Eisner ([1988], p. 7), Baxter ([1982], p. 9), Chandler ([1987], p. 58), Greiner ([1993], p. 219), Hill ([1986], p. 13, [1987], p. 9), Isihara ([1971], pp. 23--5), Jancel ([1969], pp. xix--xxii), Jellito ([1989], pp. 198--205), Khinchin ([1949], pp. 44--7), Kittel ([2004], p. 8), Landau and Lifshitz ([1980], p. 6), Lavis and Bell ([1999], p. 32), Pathria and Beale ([2011], p. 31), and Ruelle ([1969], p. 3). The principle is stated with some reluctance (which we discuss below) by Gibbs himself ([1981], p. 168), Lawden ([2005], pp. 60--1), and Tolman ([1979], pp. 62--70). The principle is implicit in Mackey ([2003], pp. 11--4), Sadovskii ([2012], pp. 5--7), and the discussion of the Gibbs formalism in Ehrenfest and Ehrenfest ([1959], pp. 47--51). A tacit appeal to the principle can be witnessed in  Feynman ([1972], Ch. 1), Huang ([1963], Chs. 7, 8), Lavis ([2015], Ch. 2), Reif ([1985], Ch. 2), Schr\"odinger ([1989], Chs. 2, 6), and Thompson ([1972], Ch. 3, [1988], Ch. 2).}  The formulations of the principle in textbooks also make it clear that AP concerns \emph{single} measurements. So AP should be read as saying that the outcome of a single measurement of property $f$ on a system in system-equilibrium is $\langle f \rangle$. The full content of AP is, however, left underdetermined until a theory of measurement is supplied. As we will see below, different schools of thought do this in different ways (associating measurement outcomes either with time averages or instantaneous values), which results in different versions of the principle.\\ 

\noindent AP lets us have the best of both worlds: one can use the formalism of ensembles to calculate equilibrium properties of a system. The importance of this principle in practical applications is beyond dispute: calculating averages is the backbone of GSM and it is what delivers the results. However, pronouncements on the status of the principle vary considerably, even among  authors who state AP explicitly. Some regard AP as the cornerstone of GSM. Chandler calls it `[t]he primary assumption of statistical mechanics' ([1987], p. 58), and Pathria and Beale regard it as the `the most important result' ([2011], p. 31). Others urge caution. Hill notes that no `completely rigorous proof is available' ([1987], p. 8), and Kittel says that `it has not been proved in general' ([2004], p. 8). Gibbs ([1981], p. 168) himself endorses the principle only for certain situations, and Lawden ([2005], p. 60) emphasises restrictions. Yet others remain silent about the question of status and scope (see, for instance, Baxter [1982], p. 9).\\ 

\noindent So there are serious questions of rationalisation. What does it mean to perform a measurement? Under what circumstances does AP apply? And how is the application justified? As previously noted, AP occupies centre stage in GSM and it is empirically successful in that it allows for the calculation of a vast array of equilibrium values that are then found to be in good agreement with experimental results. But as long as the status of a core constituent of a theory is unclear, the content of the theory itself is unclear.\\


\section{Quietism} \label{Quietism}

Mathematically minded books tend to eschew the question and simply accept the association of thermodynamic values with ensemble averages as a proposition that does not stand in need of further explanation. Thompson ([1972], p. 61, [1988], p. 30) simply equates work and pressure with ensemble averages without further comment. Similar moves can be found in Baxter ([1982], p. 9) and Feynman ([1972], pp. 1--6). This `quietist' approach is conceptually unsatisfactory because it sheds no light on the relation between GSM and TD. One simply has to accept it as a primitive assumption that the values of thermodynamic variables can be expressed in terms of ensemble averages without an explanation of why this is so. While practitioners may find it expedient to avoid the issue in this way, from a foundational point of view quietism is a deeply unsatisfactory position because it leaves the relation between GSM and thermodynamics (or indeed any macroscopic account of a system's  behaviour) unexplained.


\section{The Time Average Approach}\label{TAA}

The `standard' textbook approach explains AP by equating phase averages with time averages.\footnote{This strategy can be found, for instance, in Chandler ([1987], pp. 57--9), Hill ([1986], pp. 4--6, [1987], pp. 8--9), Isihara ([1971], pp. 24--30), Jancel ([1969], pp. xxii--xxiii), Khinchin ([1949], pp. 44--7), Kittel ([2004], pp. 7--8), and Pathria and Beale ([2011], 30--2).} The \emph{infinite time average} of $f$ is 

\begin{equation}
f^{*}(x_{0})= \lim_{\tau \rightarrow \infty} \frac{1}{\tau} \int_{t_{0}}^{t_{0}+\tau}f(\phi_{t}(x_{0})) dt, 
\end{equation}

\noindent where $x_{0}$ is the initial state of the system at time $t_{0}$. The argument runs as follows. As we have seen above, macro-variables are associated with functions $f$ on $X$. Carrying out a measurement of $f$ takes time, and hence what the measurement device registers is the time average of $f$ over the duration of the measurement. Since performing a measurement takes a long time compared to the time scale on which typical molecular processes take place, the measured result is approximately equal to the \emph{infinite} time average of the measured function. A system is {ergodic} if for all measurable functions the infinite time average is equal to the ensemble average for almost all initial conditions.\footnote{See Arnold and Avez ([1968]) for a detailed discussion of ergodicity; an intuitive introduction can be found in Frigg \emph{et al.} ([2016]).} If one now assumes that the system is ergodic, one can equate the time average and the ensemble average, which provides the sought-after connection: GSM provides phase averages which, by ergodicity, are equal to infinite time averages, and these are equal to the values obtained from measurements.\\

\noindent Implicit in this answer is the postulation that $f^{*}$ is the equilibrium value of the system. Although this is plausible given the other assumptions, there is a conceptual difference between equilibrium values and time averages and that the two are equal has to be added as an additional postulate. So we arrive at the following position. The observed value in a measurement of $f$ on a system in equilibrium is $f^{*}$. The system is ergodic and therefore it is the case that $f^{*}=\langle f \rangle$ for almost all initial conditions. One can safely neglect the `few' initial conditions for which this equality fails, and hence observed values are equal to ensemble averages, as AP has it.\\  

\noindent This argument fails for a number of reasons. First, as Malament and Zabell ([1980], pp. 342--3) and Sklar ([1973], p. 211, [1993], pp. 176--9) point out, from the fact that measurements take some time it does not follow that what is actually measured are time averages, and the association of measurement results with time averages is unjustified. Even if one could somehow argue that measurement devices do output time averages, equating these finite averages with infinite time averages is problematic. Finite and infinite averages can assume very different values even if the duration of the finite measurement is very long. This is a sticky point because the invocation of infinite time averages is crucial: if one replaces infinite with finite time averages, ergodicity no longer applies, and time averages and ensemble averages cannot be equated. Furthermore, as Earman and R\'edei ([1996]) point out, the appeal to ergodicity is by no means unproblematic. Many systems are not ergodic and if a system is not ergodic, then  equating time averages and ensemble averages is completely wrong.\footnote{One can mitigate the force of this objection somehow by requiring that the system be only $\varepsilon$-ergodic rather than `fully' ergodic (Vranas [1998]). While this resolves some problem, it does not apply to all relevant cases. For a discussion see Frigg and Werndl ([2011]).}\\

\noindent Furthermore, the assumption that $f^{*}$ is the equilibrium value of $f$ in system $S$ is introduced as primitive posit and it remains unclear why it should be adopted. And this is not only a `philosopher's worry' that practitioners can safely set aside. The function $f$ could be such that it \emph{never} assumes the time average as a value, and then it would seem rather implausible to say that $f= f^{*}$ is the system's equilibrium condition. As simple example consider the number of students registered at a university at a particular time. It may be the case that the time average of the student number is 8345.458, but there was no moment in time when 8345.458 students were registered.\\

\noindent Finally, this account makes a mystery of how we observe change. It is a simple fact that we do observe how systems approach equilibrium, and in doing so we observe macro-variables change their values. If it was true that what we observe are infinite time averages, then no change would ever be observed because these averages are by definition constant.\\

\noindent For these reasons the time average approach, at least in its current instantiation, offers no satisfactory justification of AP.


\section{Bare Probabilism}\label{Probabilism1}

An alternative approach urges a concentration on the theory's probabilistic formalism. The core of GSM is the normalised measure $\rho$, and for this reason $\rho$ has to be the centrepiece of any acceptable interpretation of GSM. We call this approach \emph{probabilism}.  As we will see, probabilism comes in a number of different variants, which differ in what events they recognise and in how much additional theoretical machinery they accept. What all variants have in common is a postulate we call \emph{$\rho$-universalism}, the assertion that $\rho$ is universally valid in that it gives the correct probabilities for \emph{all events} (recognised by the theory) to occur at \emph{all times}. Throughout this section and the next we work under the assumption that $\rho$-universalism is true. The position will be scrutinised in Section \ref{Universalism}, where we point out that $\rho$-universalism is correct only under certain conditions. \\ 

\noindent The first variant of probabilism takes the formalism of GSM at face value and interprets $\rho$ as a probability measure over the system's phase space $X$. This amounts to saying that if the ensemble is in equilibrium, then for \emph{all} regions $R$ in $X$ and for \emph{all} times $t$ the probability of the system's micro-state being in $R$ at $t$ is:

\begin{equation}\label{prob}
p(R) = \int_{R} \rho(x) dx.
\end{equation}\\

\noindent  To connect these probabilistic outcomes to observables one now adopts the notion of an \emph{instantaneous measurement}, which sees a measurement as happening at a particular instant of time. Penrose describes a measurement of that kind as `an instantaneous act, like taking a snapshot' ([2005], pp. 17--8). This idea can be made precise as follows: if a measurement is performed on a system $S$ at time $t$ and the system's microstate at time $t$ is $x$, then the measurement outcome will be $f(x)$. An obvious consequence of this definition is that measurements at different times can have different outcomes, and the values of macro-variables can change over time.\\ 

\noindent AP has not entered the scene so far, and the handling of AP is what separates what we call \emph{bare probabilism} from other forms of probabilism. Following Wallace one can insist that the quantitative content of statistical mechanics is exhausted by the statistics of observables (their expectation values, variances, etc.) and that  `statistical mechanics should not be thought of wholly or even primarily as itself a foundational project for thermodynamics' ([2015], p. 285). On such a view GSM really is just a study of the statistical properties of $\rho(x)$ with nothing else added. Hence, according to bare probabilism AP is superfluous and should be abandoned. In fact, any attempt to build a principle relating its formalism to thermodynamic equilibrium into GSM should be renounced. Instead GSM should be regarded as purely probabilistic theory. GSM characterises a system's \emph{statistical} equilibrium and specifies how likely a system will be found in certain micro-states when an instantaneous measurement is performed. That is all that there is to GSM. Any attempt to read more into GSM, in particular any attempt to read a notion of thermodynamic equilibrium into it, is misguided and should be resisted.\footnote{We have not been able to find an explicit statement of bare probabilism based on Equation (\ref{prob}) in print. However, it has been suggested to us as a natural interpretation of GSM on a number of occasions, and it sets the stage for McCoy's version of bare probabilism to which we turn shortly.}\\ 

\noindent As we have seen above, a probability density is in statistical equilibrium iff it is stationary, and on the current approach this is all that there is to say on the subject matter of equilibrium in GSM. The consequences of this brand of austerity are severe. The fact that thermodynamic equilibrium and statistical equilibrium are both equilibria does not mean that they are somehow similar, or that statistical equilibrium can serve as a stand-in for thermodynamic equilibrium when the latter is excised. In fact, statistical and thermodynamic equilibrium are not only conceptually different, the two notions also do not have the same extension. An ensemble in statistical equilibrium not only contains systems in thermodynamic equilibrium; it can also contain any number of systems that are not in thermodynamic equilibrium.\footnote{This point has been made by Uffink who emphasises that `any given system can be regarded as belonging to an infinity of different ensembles' and that it therefore `makes no sense to say whether an individual system is in statistical equilibrium or not' ([2007], p. 1005). Callender ([1999], p. 367) made the point in terms of entropy: the Gibbs entropy can assume its equilibrium value and yet a system in the ensemble can be vastly out of equilibrium. For this reason the Gibbs entropy does not reflect a system's behaviour, and the entropy of a system cannot be reduced to the Gibbs entropy.} So from the fact that an ensemble is in statistical equilibrium one cannot infer that a system randomly drawn from that ensemble is in thermodynamic equilibrium. Conversely, if a given system is in thermodynamic equilibrium, one cannot infer that this system belongs to an ensemble in statistical equilibrium. Ensemble-equilibrium and system-equilibrium remain disconnected and unmediated.\\

\noindent In a recent paper McCoy ([2018]) argues that a fundamental mistake has been made early on. The mistake is to associate GSM's probability distributions with ensembles, and thereby setting up a dichotomy between ensembles and systems that one then has to deal with through various interpretative moves. His radical remedy is to abandon ensembles altogether and regard the probability measure $\rho$ as the state of a \emph{system}. McCoy urges that `one must take seriously the idea that probability measures represent the complete physical states of individual statistical mechanical systems' (\emph{ibid}. 9), and therefore calls $\rho$ a `macrostate' (\emph{ibid}. 10). States in a statistical theory differ fundamentally from states in a deterministic theory. The state of a classical mechanical system is specified by a set of specific values of the position and momentum variables. In a statistical theory no such values can be given because the state of the system \emph{is} the probability distribution. The state of an GSM system specifies the `potentialities' of the system's observables (\emph{ibid}. 6) and not its categorical properties: the state says what properties could be observed and how likely certain outcomes are. At any given time, to specify $\rho$ is to say everything that there is to be said about the system's state at that time, and the time evolution of the system is given by specifying how $\rho$ evolves over time. We refer to this as the \emph{probabilistic states interpretation of GSM}.\\ 

\noindent The next step is to re-interpret the theory's formalism so that it provides probabilities for observables rather than micro-states. McCoy points out that this is easily done (\emph{ibid}. 13). Let $f$ be the observable of interest, and let us denote a particular value of $f$ by $F$. Then $p(F)$ is given by Equation (\ref{prob}) with $R=\{x \in X : f(x)=F\}$. The distribution $p(F)$ can now be regarded as a probability distribution for stochastically evolving observables. We now see the new picture before us: the system has macro-states $\rho$ and observables that evolve stochastically according to probabilities given by the system's macro-state. On this account there are neither ensembles nor micro-states, and so there is no bridge to gap between systems and ensembles.\\ 

\noindent One might object that $p(F)$ can be defined only with underlying micro-states and so there is no genuine stochastic dynamics for observables. McCoy anticipates this objection and submits that `one has to give up on the idea that statistical mechanical systems possess deterministically evolving classical microstates' (\emph{ibid}. 11). Taken at face value this amounts to nothing less than a rejection of atomism. McCoy is aware of the radical nature of his proposal and spends some time nuancing it. It is, however, not entirely clear what position he ends up with. On the one hand he responds `[m]y goodness, of course not!' to the charge that he repudiate the atomic hypothesis (\emph{ibid}. 12); on the other hand he submits that `people grossly overestimate the import of the alleged body of evidence for discrete, microscopic entities', which are a metaphysical posit `for which empirical evidence can hardly be decisive' (\emph{ibid}. 12).\footnote{A rejection of atomism not only raises general questions, it also precludes us, as McCoy notes, from `completing the project of the theory's founders, namely, of reducing thermodynamics to the mechanical motion of atoms and molecules' (\emph{ibid}. 5).} Eventually the position seems to be that atomism is rejected in the context of classical mechanics and that `the atomic hypothesis ultimately requires a quantum interpretation' (\emph{ibid}. 12). \\

\noindent There are serious questions about whether the rejection of the atomic hypothesis is a price worth paying (even if restricted to classical mechanics), and about whether a rejection of micro-states is as readily accommodated in quantum theory as McCoy seems to suggest. Be this as it may, our concern at this point is more basic: the problem of the relation between the two kinds of equilibrium has morphed but not disappeared. If one regards $\rho$ as the macro-state of a system, statistical equilibrium pertains to systems (and not  ensembles). But recasting statistical equilibrium as a system-equilibrium does not make it equivalent to thermodynamic equilibrium. The notions are as different as before. We have simply swapped the problem of bridging the gap between ensemble-equilibrium (defined as statistical equilibrium) and system-equilibrium (defined as thermodynamic equilibrium) for the problem of relating two different notions of system-equilibrium to one another.\\ 

\noindent At this point one might ask why such a bridge is needed at all. McCoy emphasises that `probability measures only represent the physical states of individual statistical mechanical systems \emph{qua} statistical mechanical systems' (\emph{ibid}. 3, original  emphasis). So why not simply say that there are two different perspectives on the same system --- statistical and thermodynamical --- which do not need any mediation? Qua statistical mechanical system the state of a gas is $\rho$ and it is in statistical equilibrium iff $\rho$ is stationary; qua thermodynamic system the state of the gas is given by the values of its macro-variables and it is in equilibrium if none of these variables change over time; and that is all that there is to be said about the matter.\\

\noindent This irenic perspectivalism is unconvincing, at least in the context of statistical mechanics and thermodynamics. Since McCoy renounces reductionism he may not be moved by the fact that irenic perspectivalism undercuts reduction, even though for many this would be a conclusive reason to abandon the position.  But even those who refuse to enlist in the reductionist cause are not free to simply retreat to the view that different perspectives need no mediation. Cartwright, herself a critic of reductionism, explains that there is a consistency problem ([1979]). Her main case study is Maxwell's derivation of a gas' viscosity from its mechanical properties (a case close to our current concerns!), but she illustrates the main point with the intuitive example of Eddington's two tables. Qua macro-object the table is a piece of wood that is subject to the engineering laws concerning elasticity, stress, etc. Qua micro-mechanical-object the table is a conglomerate of  atoms and molecules moving under the laws of fundamental mechanics. Cartwright then urges: `Some guarantee of consistency is required. What is to prevent macroscopic laws from moving the table one yard to the left, while all of its molecular components, following the laws of microphysics, move 30 millimeters to the right?' (\emph{ibid}. p. 85) Reductionism offers an elegant answer to the consistency problem: if one theory reduces to another theory their predictions are consistent. Those who renounce reductionist commitments are still left with the consistency problem. Irenic perspectivalism must give an account of the consistency of perspectives, and this involves giving \emph{some} account of the relation between statistical and thermodynamic equilibrium, even if it is not the kind of full-fledged reduction that the theory's founders had in mind.\\

\noindent Scientific practice poses a further challenge for irenic perspectivalism. In practice, GSM and thermodynamics often work in tandem. Calculating phase averages and relating them to equilibrium thermodynamical quantities is the bread and butter of a physicist working with GSM.\footnote{These calculations are standard and can be found in every textbook on the subject matter; see, for instance, Hill ([1987], Ch. 3), Huang ([1963], Ch. 5) and Mackey ([2003], Ch. 2).} When studying the microcanonical ensemble, the thermodynamic entropy of a system is associated with $\ln(\omega)$ and the temperature with $(d\omega/dE)^{-1}$, where $\omega$ is the integral of $\rho$ over a thin sheet around the energy hypersurface of the system. When working with the canonical ensemble it is standard practice to calculate the Helmholtz free energy $F$ as function of $\rho$:  $F=-kT \log Z$, where $Z$ is the partition function, which is essentially the normalisation factor of $\rho$. These quantities are then inserted into the equations of thermodynamics and used for calculations. In many cases even the thermodynamic equations of state are derived from GSM. This practice must appear mysterious from a point of view that sees GSM as completely disconnected from thermodynamics.\\

\noindent Stated without further qualifications, bare probabilism is too weak an interpretation of GSM. A promising way to qualify the position is to study fluctuations. We introduce fluctuations in the next section and then ask what their status is vis-\`{a}-vis bare probabilism.


\section{Fluctuations}\label{Fluctuations}

A way to answer the questions left open by bare probabilism is to study fluctuations. Since measurement is assumed to be instantaneous, the system's instantaneous state $x(t)$ is allowed to fluctuate away from $\langle f \rangle$. One can now ask how far from $\langle f \rangle$ the system is. The difference between $\langle f \rangle$ and the instantaneous value of $f$ at time $t$ is a \emph{fluctuation}:

\begin{equation}\label{time-fluctuation}
\Delta(t) = f(x(t)) - \langle f \rangle.
\end{equation}

\noindent We call $| \Delta(t) |$ the \emph{magnitude} of the fluctuation.\\  

\noindent Recall that $\rho$-universalism is the position that $\rho$ provides the correct probabilities for a system's state to be in region $R$ at time $t$ for all $R$ in $X$ and for all $t$. Under that assumption, Equation (\ref{prob}) can be used to calculate the probability of fluctuations of a certain size.\footnote{We use the standard version of probabilism and assume that $\rho$ provides probabilities for the system to be in certain sets of micro-states. This is choice of convenience; all points that follow could equally be made in McCoy's probabilistic state interpretation.} One finds that at any given time $t$ the probability that the magnitude of $\Delta(t)$ lies in the interval $\delta := [\delta_{1}, \delta_{2}]$, where $\delta_{1}$ and $\delta_{2}$ are real numbers such that $0 \leq \delta_{1} \leq \delta_{2}$, is:

\begin{equation}\label{fluc-prob}
p(\delta) = \int_{D} \rho(x) dx, 
\end{equation}

\noindent where $D = \{x \in X \mid  \delta_{1} \leq | \Delta(t) | \leq \delta_{2} \}$. This equation informs us about the probability that the system exhibits fluctuations of a certain magnitude at a certain time $t$ (under the assumption of $\rho$-universalism).\\ 

\noindent The leading idea now is to use the probabilities of fluctuations to characterise a system's macroscopic behaviour. As we have seen in Section \ref{Ensembles}, a system is in thermodynamic equilibrium iff all change has come to a halt and the system's macro-variables assume constant values. By definition, these constant values are the thermodynamic equilibrium values. Expressing this condition in terms of fluctuations yields that a system is in thermodynamic  equilibrium iff there are no fluctuations at all. In this case $\langle f \rangle$ is the thermodynamic equilibrium value, and AP is true. However, statistical mechanical systems do not meet this condition and so requiring the complete absence of fluctuations is a dead end.\footnote{The source of the problem is the fact that measure-preserving dynamical systems exhibit Poincar\'{e} recurrence and time-reversal invariance (Callender [2001]).} Fluctuations are inevitable and a criterion that has any chance of being met can only require that fluctuations are somehow unlikely. \\

\noindent A plausible candidate for such a condition is the requirement that small fluctuations are likely while medium size and large fluctuations are unlikely. In formal terms, this amounts to requiring that $p(\delta)$ is noticeably different from zero only for intervals that lie close to zero and becomes vanishingly small for intervals far away from zero. A system that exhibits this pattern approximately mimics thermodynamic behaviour, so we refer to fluctuations that have these characteristics as \emph{thermodynamic fluctuations}. In fact, fluctuations being thermodynamic furnishes an approximate justification of AP because the value of $f$ for a system in equilibrium is approximately constant and close to $\langle f \rangle$. Following Callander's ([2001]) sage advice of not taking thermodynamics too seriously, one can accept an approximate justification of AP as sufficient. In this vein Hill submits that the validity of the identification of observable values with ensemble averages is legitimate only when the fluctuations of $f$ away from $\langle f \rangle$ are small ([1987], pp. 9--10), and Schr\"odinger declares ([1989], p. 35) that `[m]ean value, most probable value, any values that occur with non-vanishing probability --- all become the same thing'.\\

\noindent It cannot be taken for granted that fluctuations are thermodynamic. Whether the fluctuations in a given situation exhibit this pattern depends both on the system's dynamics and the macro-variables. While the fluctuations in some applications of GSM turn out to be thermodynamic as expected, there are systems in which large fluctuations turn out to be more likely than small fluctuations. Fluctuations in the Kac-ring with the standard macro-state structure of the number of black and white balls as discussed by Lavis ([2008]) are not thermodynamic. In our ([2017a]) we also discuss two models in which fluctuations are not thermodynamic. The first is the baker's gas with a macro-variable $V$ that takes integer values on cells of the standard Boltzmannian partition; the second is the finite Ising model for certain termperature values with the internal energy as a macro-variable. And in our ([unpublished]) we show that the same result also occurs in the six-vertex model for certain temperature values. That fluctuations are thermodynamic has to be ascertained on case-by-case basis.\\

\noindent What is the status of AP vis-\`{a}-vis bare probabilism? There are two options. As we have seen above, bare probabilism insists that the quantitative content of statistical mechanics is exhausted by the statistics of observables and that nothing else should be added to it. This means that AP has no place \emph{in} the theory and has to be located somewhere else. The question is where. A plausible answer is to interpret AP as a \emph{bridge law} that is invoked when GSM is connected to TD. Bridge laws relate two different theories to each other by establishing a relation between the terms of the theories, typically with the aim of reducing one theory to the other.\footnote{For a discussion of reductionism and bridge laws see Dizadji-Bahmani \emph{et al.} ([2010]).} A classical example of a  bridge law is the law in the kinetic theory of gases which says that temperature is proportional to the mean kinetic energy of the molecules, thereby establishing a link between a mechanical concept (kinetic energy) and a thermodynamic quantity (temperature). AP can be thought of as a bridge law of that kind. While not being part of GSM proper, it then nevertheless establishes a link between concepts in GSM and thermodynamics.\\ 

\noindent This is a workable position, but it is important to get clear on the commitments that this implies. Locating AP outside GSM might suggest that it is a `voluntary addition' that can be invoked when it is convenient, but ignored otherwise. Those who reject an understanding of SM as a foundational project for thermodynamics would then be free to ignore AP altogether. Unfortunately there is rather less freedom of choice. As noted previously, it is common practice in applications to calculate a certain quantity using the averaging methods of GSM and then plug it into thermodynamic equations. For this to be legitimate, the requisite bridge law, namely AP, has to hold. So bare probabilism must ensure that AP holds whenever GSM is used in tandem with TD. Adding this proviso to bare probabilism gives \emph{qualified probabilism}.\\

\noindent An alternative is to take statements in textbooks at face value and regard AP, in Chandler's words, as a primary assumption of statistical mechanics (see Section \ref{Ensembles}). On such an interpretation AP is an integral part of GSM. We call this position \emph{fluctuation probabilism}. Fluctuation probabilism has to restrict dynamical laws and observables that are allowable in GSM to those that produce thermodynamic fluctuations. Restrictions of this kind are common: fields in electrodynamics must be twice differentiable and forces in Newtonian mechanics must be Lipschitz continuous for unique solutions to exist. However, as we will see later, it is less than clear what conditions a system must satisfy for fluctuations to be thermodynamic and so making this condition part of GSM introduces an certain degree of indeterminacy about the theory's content.\\ 

\noindent However, at the end of the day the choice between fluctuation probabilism and qualified probabilism is an aesthetic matter. Both accounts have to come to terms with fluctuations; they differ only in where they locate this problem. The main challenge for both approaches is that they are based on $\rho$-universalism. In the next section we show that $\rho$-universalism is false and that $\rho$ gives correct probabilities for events to occur at all times only if the system satisfies certain conditions.


\section{Scrutinising $\rho$-Universalism}\label{Universalism}

All versions of probabilism presuppose $\rho$-universalism. Unfortunately, that doctrine is not true in general and $\rho$ can be used to calculate probabilities of events only under certain circumstances (that is, Equations (\ref{prob}) and (\ref{fluc-prob}) have restricted validity). The nature of these restrictions depends on how fluctuations are interpreted. In this section we discuss two interpretations and spell out their implications.\\ 

\noindent On the first interpretation we consider the fluctuations that arise in the \emph{same} system when we observe its behaviour over time. This amounts to tracking a system over an extended period of time (usually one considers the limit $t \rightarrow \infty$) when the system starts in a particular initial condition $x_{0}$ at time $t=0$ and its state  evolves under $\phi_{t}$. The system's instantaneous state at any time $t$ then is $x(t) = \phi_{t}(x_{0})$. $\rho$-universalism then asserts that for all $R \subset X$ and all times $t$, Equation (\ref{prob}) gives the long-run fraction of time that $x(t)$ is in $R$ as time goes to infinity and Equation (\ref{fluc-prob}) gives the correct probability for fluctuations of certain magnitude to occur.\\ 

\noindent Unfortunately this is not the case. The problem is one that has been described by Lawden ([2005], pp. 60--1) and Sklar ([1993], pp. 193--4). The dynamics of the system can have conserved quantities. In that case the state space $X$ decomposes into invariant subsets, i.e. sets that are mapped onto themselves under the dynamics of the system. A trajectory then remains confined to the subset in which its initial condition lies and it never wanders off into other invariant subsets. This, however, invalidates Equation (\ref{prob}), and with it Equation (\ref{fluc-prob}). To see how this happens assume that $X$ decomposes into two invariant subsets $X_{1}$ and $X_{2}$, and that $\rho(X_{1})\neq 0$ and $\rho(X_{2})\neq 0$.\footnote{The argument generalises trivially to any number of subsets. If the distribution $\rho$ vanishes on the invariant subsets, a similar argument can be made by choosing $X_{1}$ (and $X_{2}$) to be a collection of invariant subsets with $\rho(X_{1})\neq 0$ (and $\rho(X_{2})\neq 0$).} Assume that the initial condition $x_{0}$ of the system whose behaviour we are tracing over time lies in $X_{1}$. Because $X_{1}$ is an invariant set, the system will then never leave $X_{1}$ and the probability of the system's state ever being in $X_{2}$ is zero. Yet $\rho$ assigns a non-zero probability to the system being in $X_{2}$. So $\rho$ makes wrong predictions. As a result also the predictions for fluctuations can be wrong. The function $f$ may be such that it assumes certain values only in $X_{2}$ and hence fluctuations away from the mean of a certain magnitude only occur when the state of the system is in $X_{2}$. Equation (\ref{fluc-prob}) assigns a non-zero probability to such fluctuations and yet the system never exhibits them.\footnote{A referee suggested that this problem has an `easy and obvious remedy', namely to `restrict the probability distribution to the relevant invariant subset that describes the system with a particular value of the conserved quantity'. Such restrictions would lead to distributions that are very different from the ones usually considered in GSM and, as far as we can tell, such restrictions are not carried out in the practice of GSM. One reason for this is that the invariant subsets are usually not known, which makes it impossible to specify a suitably restricted probability distribution. Furthermore, invariant sets can be of measure zero, and then the restriction procedure provides no meaningful results.}\\

\noindent One way of avoiding these difficulties is to require that the system's dynamics is such that it can access all parts of $X$. In deterministic systems this is tantamount to requiring that the system is ergodic.\footnote{For Markov processes (the kind of stochastic dynamics that is usually assumed in the context of stochastic GSM) it amounts to requiring that the Markov process is irreducible. The six-vertex model, for instance, is an irreducible Markov process.} Some systems are of this kind: for all we know the hard ball gas is ergodic. But other systems violate these conditions. The Kac ring is a case in point. The uniform distribution on its state space is a Gibbsian equilibrium, and yet its state space has an ergodic decomposition, meaning that it consists of a number of invariant subsets (Lavis [2005]). This is the situation we described in the previous paragraph, and all the problems we mentioned arise in the Kac ring. Another example where the same problem arises is a gas of non-interacting particles in a multi-mushroom-box (Werndl and Frigg [forthcoming]). Furthermore, although rigorous proofs are currently out of reach, it is generally expected that many fluids and solids will not be ergodic, which, again, leads to the same difficulties (Uffink [2007], p. 1017).\\

\noindent A possible response is that requiring ergodicity is asking for too much. All that is needed is that if a system is `trapped' in a certain part of state space, this does not influence the probabilities of fluctuations. This happens just in case the proportion of states for which $f$ assumes a particular value is the same in each invariant subset. In this case the $p(\delta)$ are the correct probabilities even if the system cannot access the full state space. In other words, what is needed is not full ergodicity but the weaker condition that the invariant sets of the dynamics (if any) and the function $f$ are attuned so as to `mask' the fact that the system is trapped in certain invariant sets. We call this the \emph{masking condition}. Under what circumstances does the masking condition hold? As far as we can see, there are no general criteria to decide whether a system has this property. In fact, whether a trapping will be masked in this way depends both on the time evolution $\phi_{t}$ and the macro-variable $f$, and one will have to look at each $f-\phi_{t}$ pair to come to a verdict. A positive result can in no way be taken for granted. If a system is not ergodic, there will always be many macro-variables for which the masking condition fails.\\

\noindent The root of the difficulties we encountered so far (and which led to the formulation of the masking condition) was the adoption of an interpretation of $\Delta(t)$ as the fluctuation one observes on a \emph{single system} when one traces the system's state over time. In the light of these difficulties one might deem this an unsuitable choice and suggest that an alternative notion of fluctuation should be used. This alternative interpretation emerges from Gibbs' original description of an ensemble:

\begin{quote}
`We may imagine a great number of systems of the same nature [...] And here we may set the problem, not to follow a particular system through its succession of configurations, but to determine how the whole number of systems will be distributed among the various conceivable configurations and velocities at any required time [...]' (Gibbs [1981], p. v)
\end{quote}

\noindent This suggests thinking of an ensemble like an urn of balls. An ensemble is a large collection of (imaginary) systems and making an observation amounts to drawing one system out of the ensemble. The distribution $\rho$ then specifies the probability for drawing a system whose state $x$ lies in a certain part of the state space in much the same way in which the fraction of red balls in the urn specifies the probability of drawing a red ball. If the drawn system is in state $x$, the observed value of $f$ is $f(x)$, and hence $\rho$ also specifies the probabilities for certain values of $f$.\\

\noindent Gibbs relied on this idea when he observed that `[w]hat we know about a body can generally be described most accurately and most simply by saying that it is one taken at random from a great number (ensemble) of bodies which are completely described' ([1981], p. 163). This way of thinking about ensembles is widespread in the physics literature. Hill says that `if a system is chosen at random from the ensemble at time $t$, the probability of finding the phase point representative of its dynamical state in $dp \, dq$ about the point $p,q$ is $f \, dp \, dq$' ([1987], pp. 4--5), where $f$ stands for the Gibbsian $\rho$. Tolman describes the density $\rho$ as giving the probability that `the phase point for a system chosen at random from the ensemble would be found at time $t$ to have the specified values of the $q$'s and $p$'s' ([1979], p. 47). Lawden states that `the frequency with which a particular state is found in the ensemble is proportional to its associated probability [i.e.\ $\rho$]' ([2005], pp. 56--57). Agarwal and Eisner ([1988], p. 7), Isihara ([1971], p. 21), Jellito ([1989], pp. 192--3), Penrose ([2005], pp. 97--8), Reif ([1985], pp. 47--55), Schr\"odinger ([1989], p. 9), and Thompson ([1972], p. 56) make statements to the same effect.\\

\noindent If one understands $\rho$ as the probability of finding a system randomly drawn from the ensemble in state $x$,  Equation (\ref{prob}) is true by definition. One can then interpret the $f(x(t))$ in the definition of $\Delta(t)$ as the result one gets when drawing a system from the ensemble at random at time $t$ (rather than as the observation of $f$ on the same system at different times). Under this interpretation $\Delta(t)$ is the statistical fluctuation one finds in consecutive random draws. Since consecutive draws are by assumption independent from one another, they do not depend on initial conditions. Hence the problem with invariant sets vanishes and Equation (\ref{fluc-prob}) is true under all circumstances.\\ 

\noindent This is an elegant move, but it raises a serious question. In laboratory experiments we observe the \emph{same} system at consecutive times rather than drawing different systems out of an urn at random. This difference is significant because successive observations on the same system are generally not independent in the way in which draws from an urn are, and so it remains unclear what we can infer about laboratory experiments from GSM under an `urn interpretation'. \\ 

\noindent One could respond to this worry by emphasising the qualification that observations on the same system are \emph{generally} not independent and argue that the systems of interest in GSM fall into a special class of system for which independence holds. The systems to which GSM applies, so the argument goes, are such that they effectively re-randomise between measurements. Kittel discusses this issue and expresses confidence that this happens in the sort of systems we are interested in: 

\begin{quote}
`the complex systems with which we are dealing appear to randomize themselves between observations, provided only that the observations follow each other by a time interval longer than a certain characteristic time called the relaxation time. The relaxation time describes approximately the time required for a fluctuation (spontaneous or arranged) in he properties of a system to damp out.' (Kittel [2004], p. 7)
\end{quote}

\noindent How can one characterise systems that have this property? The randomization that is needed to get the position off the ground is of the following kind. Let $A$ and $B$ be two arbitrary subsets of $X$. We write `$A_{t}$' to indicate that the state of the system that was drawn out of the urn at time $t$ was in $A$, and likewise for `$B_{t}$'. The independence condition needed for the urn interpretation to be correct then is this: there exists a timespan $\tau$ (the relaxation time) so that for all $A, B \subseteq X$ we have $p(B_{t_{2}}|A_{t_{1}}) = p(B_{t_{2}})$ provided $t_{2}-t_{1}\geq \tau$. That is, draws at later times have to be probabilistically independent of draws at earlier times provided that the distance between them is more than the relaxation time (which is presumed to be relatively short).\\ 

\noindent This condition can be true only if the system is mixing, and even then the randomization can be achieved only in the infinite limit and never over finite times.\footnote{See Berkovitz \emph{et al.} ([2006]) and  Werndl ([2009]).} Moreover, mixing implies ergodicity and, as we have already seen, many systems in statistical mechanics are not ergodic. A fortiori they are not mixing and therefore do not randomize in the required manner. On might reply that we have required too much by stipulating that $p(B_{t_{2}}|A_{t_{1}}) = p(B_{t_{2}})$ has to hold \emph{for all} subsets $A$ and $B$ of $X$. One can prove that mixing is equivalent to the condition that for \emph{all} functions $f \in L^2(X)$ (where $L^2(X)$ is the space of square integrable functions on $X$) the sequence of $\int_{X}f(\phi_{t}(x))d\mu$ converges to $\int_{X}f(x)d\mu$ as $t$ tends towards infinity (Katok and Hasselblatt [1993], p. 152). This shows that requiring that the system is mixing is tantamount to requiring that independence holds with respect to \emph{all}  functions $f$. But this may well be too strong. What is needed to make GSM work is only the weaker assumption that independence holds with respect to certain \emph{selected} functions that are deemed relevant in a given context. Let $f$ be a function on $X$. We then say that a system is \emph{$f$-independent} iff $\int_{X}f(\phi_{t}(x))d\mu$ converges to $\int_{X}f(x)d\mu$ as $t$ tends towards infinity. The relevant concept then is $f$-independence rather than mixing. Furthermore, one might argue that it is unnecessary to require that complete independence is reached; all that is needed is that correlations relax below a level where they become negligible. So it suffices to require that systems are approximately $f$-independent for certain selected variables $f$.\\

\noindent Restricting the application of GSM to systems that are  approximately $f$-independent for a few selected variables is a successful move as far as it goes, but how far is this? The problem is that reaching approximate independence in finite time is a relational property that a macro-variable $f$ possesses with respect to a particular time evolution $\phi_{t}$, and whether or not independence holds depends on both $f$ and $\phi_{t}$. On this approach one would have to check every  $f-\phi_{t}$ pair before being licensed to apply GSM.\\

\noindent A radical way around this problem is to try to vary the setup and endeavour to produce an ensemble empirically, for instance by destroying the system after every measurement and recreate it in a new state (or, less dramatically, by interrupting its time evolution and resetting its state). In this way one could, at least in principle, produce a large number of independent systems that, taken together, approximate a Gibbsian ensemble.\\ 

\noindent There are doubts that this will work.\footnote{Furthermore, this project is of questionable legitimacy because in doing so one gives up the aim of describing the evolution of single systems, which is important in statistical mechanics.} The systems produced in this way would approximate the Gibbsian ensemble only if the process of ensemble preparation was such that the systems ended up being produced according to the measure of the ensemble. As Leeds ([1989], pp. 329--33) and Werndl ([2013], pp. 473--6) point out, there is no reason to believe that the repeated preparation of a system in a certain macro-condition will be such that the collection of all systems satisfies the relevant Gibbsian distribution (e.g. the microcanonical distribution). In fact, depending on the details of the process (who prepares the system and how) different distributions could be obtained. The best one can hope for is that the resulting distribution is absolutely continuous with $\rho$  (i.e. it agrees with $\rho$ on all sets of measure zero), but this allows for empirical outcomes that diverge significantly from $\rho$.\\

\noindent There is a temptation to reply that this is an artefact of the state preparation and that the time evolution of the system irons out the discrepancies between the distribution resulting from the state preparation and the ensemble distribution. So rather than saying that one observes a certain distribution $\rho$ when repeatedly preparing a system, one should be committed to the claim that one finds $\rho$ when repeatedly preparing a system \emph{and} letting it run for a fixed time $t$ (which is long enough for the system to `settle down'). Unfortunately, this does not solve the problem either, at least not in general. For it to be the case that all initial distributions\footnote{All distributions that are absolutely continuous w.r.t. the Lebesgue measure.} converge to the ensemble distribution, the dynamics has to be mixing (Werndl [2009]), which gets us back to all the difficulties with mixing that we have already discussed.\\ 

\noindent In sum, the Gibbsian $\rho$ can be used to calculate correct fluctuation probabilities only if the time evolution $\phi_{t}$ and the macro-variable $f$ work in tandem to guarantee that the masking condition is satisfied or that $f$-independence holds at least approximately. If one of those conditions holds, then the fluctuation probabilities as given in Equation (\ref{fluc-prob}) truthfully reflect a system's dynamics and can be used to assess its equilibrium behaviour.\\ 

\noindent Unfortunately, both conditions are strong and cannot be taken for granted. And the problem is not one that can be brushed aside as `irrelevant for all practical purposes'. The possibility of these conditions not holding is not merely a remote mathematical possibility that is inconsequential for applications; there are examples where these conditions \emph{do} fail. An example where the masking condition fails is the Kac ring. Lavis ([2008], p. 686) shows that there are solutions which stay only 46.58\% of their time in a certain region $X_{1}$ of the state space (the Boltzmannian equilibrium region) where Equation (\ref{prob}) would predict that it should stay in that region 99.999\% percent of the time. If one now chooses an $f$ so that it assumes one value in $X_{1}$ and another value in the rest of the phase space, then the masking condition is violated. The condition also fails in a gas of non-interacting particles in a multi-mushroom box with a macro-variable indicating whether the solution moves back and forth between the first and second mushroom or between the second and third mushroom (Werndl and Frigg [forthcoming]). These systems are not ergodic, and so one would expect such failures. Since other systems are likely to be in the same category, one would expect similar failures in other systems too. \\

\noindent The condition of approximate $f$-independence is violated whenever a system with a continuous dynamics is observed in intervals that are so short that no independence is achieved, not even approximately. The issue is that if a system is continuous, the spreading out of regions in phase space happens continuously and hence independence can be achieved only after a sufficiently long time period has passed. This is not just a theoretical possibility. In fact, relaxation times can be observed experimentally, which involves observing the system in intervals that are shorter than the time needed for them to randomize. In such a situation Gibbsian probabilities will not be correct.\\

\noindent There is a legitimate question how much weight these examples bear. Due to the mathematical complexities of the systems usually studied in GSM, explicit results are available only for simple systems like the Kac ring and the mushroom gas. Whether, and under what circumstance, masking and approximate $f$-independence hold in more realistic systems is an interesting question that deserves further attention. And the question is not merely one of doing `routine background checks' that one confidently expects to return the right verdict. As we noted above, both conditions are highly non-trivial and we should expect them to fail in some systems. Indeed, in some systems both conditions fail. This happens, for instance, in the KAC ring when periods of time are considered that are shorter than the relaxation time (and we associate the system's thermodynamic equilibrium values with the values of $f$ that the system assumes most of the time). Surprisingly, GSM and AP still work in this particular case even though probabilism (in any of its forms) fails, and hence offers no justification for AP. This suggests that probabilism does not provide a satisfactory \textit{general} interpretation of GSM.\footnote{For our explanation of why AP works in these cases see our ([unpublished])}\\


\section{Conclusion}\label{Conclusion}

We reviewed different interpretations of GSM and came to the conclusion that among the currently available interpretations qualified probabilism and fluctuation probabilism are the most promising options. However, in their original formulation they were based on $\rho$-universalism, which is false. The measure $\rho$ gives correct probabilities for events to occur at all times only if the system under study satisfies the masking condition or the $f$-independence condition. So probabilism (of either variety) is true only if at least one of these conditions holds.\\ 

\noindent If this is the case, GSM allows us to calculate the probability of fluctuations. But, as we have seen above, this does not \emph{ipso facto} imply that fluctuations turn out to be thermodynamic. The account can give the right fluctuations probabilities, but the fluctuations can be of the wrong kind, for instance in that large fluctuations can turn out to be more likely than small fluctuations. This is not merely a remote conceptual possibility. As noted in Section \ref{Fluctuations}, there are systems in which fluctuations are not thermodynamic and in which AP fails.\\ 

\noindent Finally, it is crucial to note that not all justifications of AP have to be given in the framework of the fluctuation approach. There is a temptation to regard the fluctuation approach as the `natural' interpretation of GSM, and then discuss the validity of AP only \emph{within} the fluctuation approach (where the question comes down to whether fluctuations are thermodynamic). This temptation must be resisted. There are cases where AP holds and the fluctuation approach fails, and narrowing our focus on the fluctuation account makes us blind for such cases. The Kac ring with the standard magnetisation macro variable is a case in point. As we have seen above, the Kac ring satisfies neither the masking condition (for short time periods) nor $f$-independence and yet it turns out that $\langle f \rangle$ is equal to the equilibrium value. Intuitively this is the case because even though the various regions where $f$ takes values deviating considerably from the equilibrium value are distributed symmetrically over the state space and hence `cancel out' when the average is calculated. Hence, if we are interested in justifying AP, we must sometimes look beyond fluctuations, and indeed beyond probabilism.\\


\section*{Acknowledgments}

This paper grew out of a discussions we had with Wayne Myrvold. We would like to thank Wayne for sharing his knowledge with us for giving so generously of his time. We also want to thank Erik Curiel, Casey McCoy, Sam Fletcher, David Lavis, Joshua Luczak, James Wills and two anonymous referees for helpful comments on earlier versions. We presented the paper in Munich, Irvine and Salzburg and would like to thank the audiences for engaging discussions. 


\section*{References}

\begin{list}{}{    \setlength{\labelwidth}{0pt}
    \setlength{\labelsep}{0pt}
    \setlength{\leftmargin}{24pt}
    \setlength{\itemindent}{-24pt}
  }

\item Agarwal, B. K. and Eisner, M. [1988]: \emph{Statistical Mechanics}, 2nd ed., New Delhi: New Age International.

\item Arnold, V. I. and Avez, A. [1968]: \emph{Problems of Classical Mechanics}, New York and Amsterdam: W. A. Benjamin.

\item Baxter, R. J. [1982]: \emph{Exactly Solved Models in Statistical Mechanics}, London: Academic Press.
 
\item Berkovitz, J., Frigg, R. and Kronz, F. [2006]: `The Ergodic Hierarchy, Randomness and Hamiltonian Chaos', \emph{Studies in History and Philosophy of Modern Physics}, \textbf{37}, pp. 661--91.

\item Buchdahl, H. A. [1975]:  \emph{Twenty Lectures on Thermodynamics}, Rushcutters Bay: Pergamon Press.

\item Callender, C. [1999]: `Reducing Thermodynamics to Statistical Mechanics: The Case of Entropy', \emph{Journal of Philosophy}, \textbf{96}, pp. 348--73.

\item Callender, C. [2001]: `Taking Thermodynamics Too Seriously',  \emph{Studies in the History and Philosophy of Modern Physics}, \textbf{32}, pp. 539--53.

\item Chandler, D. [1987]: \emph{Introduction to Modern Statistical Mechanics}, New York: Oxford University Press.

\item Cartwright, N. [1979]: `Do Token-Token Identity Theories Show Why We Don't Need Reductionism?', \emph{Philosophical Studies}, \textbf{36}, pp. 85--90.

\item Dizadji-Bahmani, F., Frigg, R. and Hartmann, S. [2010]: `Who's Afraid of Nagelian Reduction?', \emph{Erkenntnis}, \textbf{73}, pp. 393--412.

\item Earman, J. and R\'edei, M. [1996]: `Why Ergodic Theory Does Not Explain the Success of Equilibrium Statistical Mechanics', \emph{The British Journal for the Philosophy of Science}, \textbf{47}, pp. 63--78.

\item Ehrenfest, P. and Ehrenfest, T. [1959]: \emph{The Conceptual Foundations of the Statistical Approach in Mechanics}, Mineola: Dover Publications. 

\item Fermi, E. [2000]: \emph{Thermodynamics}, Mineola/NY: Dover Publications.

\item Feynman, R. P. [1972]: \emph{Statistical Mechanics: A Set of Lectures}, Reading, MA: W.A. Benjamin.

\item Frigg, R.,  Berkovitz, J. and Kronz, F. [2016]: `The Ergodic Hierarchy', in E. Zalta (\emph{ed.}), \emph{The Stanford Encyclopedia of Philosophy},  \textless{}plato.stanford.edu/\newline{archives/sum2016/entries/ergodic-hierarchy/\textgreater.}

\item Frigg, R. and Nguyen, J. [2018]: `Mathematics is Not the Only Language in the Book of Nature',  \emph{Synthese}, Online First, \textless{doi}.org/10.1007/s11229-017-1526-5\textgreater. 

\item Frigg, R. and Werndl, C. [2011]: `Explaining Thermodynamic-Like Behaviour in Terms of Epsilon-Ergodicity', \emph{Philosophy of Science}, \textbf{78}, pp. 628--52.

\item Frigg, R. and Werndl, C.  [forthcoming]: `Equilibrium in Gibbsian Statistical Mechanics', forthcoming in E. Knox and A. Wilson (\emph{eds}), \emph{Routledge Companion to Philosophy of Physics}.

\item Gibbs, J. W. [1981]: \emph{Elementary Principles in Statistical Mechanics}, Woodbridge: Ox Bow Press.

\item Greiner, W.,  Neise, L. and  St\"ocker, H. [1993]: \emph{Thermodynamik und statistische Mechanik. Ein Lehr- und \"Ubungsbuch}, 2nd ed., Thun and Frankfurt am Main: Harri Deutsch.

\item Guggenheim, E. A. [1967]: \emph{Thermodynamics: An Advance Treatment for Chemists and Physicists}, Amsterdam: North-Holland.

\item Gyftopoulos, E. P. and Beretta, G. P. [2005]: \emph{Thermodynamics: Foundations and Applications}, Mineola/NY: Dover Publications.

\item Hill, T. L. [1986]: \emph{An Introduction to Statistical Thermodynamics}, Mineola, NY: Dover Publications.

\item Hill, T. L. [1987]: \emph{Statistical Mechanics: Principles and Selected Applications}, Mineola, NY: Dover Publications.

\item Honig, J. [1999]: \emph{Thermodynamics}, San Diego: Academic Press.

\item Huang, K. [1963]: \emph{Statistical Mechanics}, New York, London, and Sydney: John Wiley \& Sons.

\item Isihara, A. [1971]: \emph{Statistical Physics}, London: Academic Press.

\item Jancel, R. [1969]: \emph{Foundations of Classical and Quantum Statistical Mechanics}, Oxford: Pergamon Press.

\item Jelitto, R. J. [1989]: \emph{Theoretische Physik 6: Thermodynamik und Statistik}, 2nd ed., Wiesbaden: AULA-Verlag.

\item Katok, A. and Hasselblatt, B. [1993]: \emph{Introduction to the Modern Theory of Dynamical Systems}, Cambridge: Cambridge University Press.

\item Khinchin, A. I. [1949]: \emph{Mathematical Foundations of Statistical Mechanics}, Mineola, NY: Dover Publications.  

\item Kittel, C. [2004]: \emph{Elementary Statistical Physics}, Mineola, NY: Dover Publications.

\item Kubo, R. [1968]: \emph{Thermodynamics. An Advance Course with Problems and Solutions}, Amsterdam: North-Holland.

\item Landau, L. D. and  Lifshitz, E. M. [1980]: \emph{Statistical Physics}, Oxford: Elsevier.

\item Lavis, D. A. [2005]: `Boltzmann and Gibbs: An attempted reconciliation', \emph{Studies in History and Philosophy of Modern Physics}, \textbf{36}, pp. 245--73.

\item Lavis, D. A. [2008]: `Boltzmann, Gibbs and the Concept of Equilibrium', \emph{Philosophy of Science}, \textbf{75}, pp. 682--96.

\item Lavis, D. A. [2015]: \emph{Equilibrium Statistical Mechanics of Lattice Models}, Dordrecht, Heidelberg, New York, and London: Springer.

\item Lavis, D. A. and Bell, G. M. [1999]: \emph{Statistical Mechanics of Lattice Systems Volume 1: Closed-Form and Exact Solutions}, Berlin, Heidelberg, and New York: Springer.

\item Lawden, D. F. [2005]: \emph{Principles of Thermodynamics and Statistical Mechanics}, Mineola, NY: Dover Publications.

\item Leeds, S. [1989]: `Malament and Zabell on Gibbs phase averaging', \emph{Philosophy of Science}, \textbf{56}, pp. 325--40.

\item Mackey, M. C. [2003]: \emph{Time's Arrow: The Origins of Thermodynamic Behavior}, Mineola, NY: Dover Publications.

\item Malament, D. and Zabell, S. L. [1980]: `Why Gibbs Phase Averages Work --- The Role of Ergodic Theory', \emph{Philosophy of Science}, \textbf{47}, pp. 339--49.

\item McCoy, C. D.  [2018]. `An Alternative Interpretation of Statistical Mechanics', \emph{Erkenntnis}. Online First, \textless{}doi.org/10.1007/s10670-018-0015-7\textgreater.

\item Myrvold, W. C. [2016]: `Probabilities in Statistical Mechanics', in A. H\'ajek and C. Hitchcock (\emph{eds}), \emph{The Oxford Handbook of Probability and Philosophy}, Oxford: Oxford University Press, pp. 573--600.

\item Pathria, R. K. and Beale, P. D. [2011]: \emph{Statistical Mechanics}, 3rd ed., Oxford: Elsevier.

\item Penrose, O. [2005]: \emph{Foundations of Statistical Mechanics: A Deductive Treatment}, Mineola, NY: Dover Publications.

\item Pippard, A. B. [1966]: \emph{Elements of Classical Thermodynamics}, Cambridge: Cambridge University Press.

\item Reif, F. [1985]: \emph{Fundamentals of Statistical and Thermal Physics}, Singapore: McGraw-Hill.

\item Reiss, H. [1996]: \emph{Methods of Thermodynamics}, Mineola, NY: Dover.

\item Ruelle, D. [1969]: \emph{Statistical Mechanics: Rigorous Results}, London: Imperial College Press.
 
 \item Sadovskii, M. V. [2012]: \emph{Statistical Physics}, Berlin: De Gruyter.
 
\item Schr\"odinger, E. [1989]: \emph{Statistical Thermodynamics}, Mineola, NY: Dover Publications.

\item Sklar, L. [1973]: `Statistical Explanation and Ergodic Theory', \emph{Philosophy of Science}, \textbf{40}, pp. 194--212.

\item Sklar, L. [1993]: \emph{Physics and Chance. Philosophical Issues in the Foundations of Statistical Mechanics}, Cambridge: Cambridge University Press. 

\item Thompson, C. J. [1972]: \emph{Mathematical Statistical Mechanics}, New York: Macmillan.

\item Thompson, C. J. [1988]: \emph{Classical Equilibrium Statistical Mechanics}, Oxford: Clarendon Press.

\item Tolman, R. C. [1979]: \emph{The Principles of Statistical Mechanics}, Mineola, NY: Dover Publications.

\item Uffink, J. [2007]: `Compendium of the foundations of classical statistical physics', in J. Butterfield and J. Earman (\emph{eds}), \emph{Philosophy of Physics},  Amsterdam: North Holland, pp. 923--1047.

\item van Lith, J. [1999]: `Reconsidering the Concept of Equilibrium in Classical Statistical Mechanics', \emph{Philosophy of Science}, \textbf{66} (Supplement), pp. 107--18.

\item van Ness, H. C. [1969]: \emph{Understanding Thermodynamics}, New York: Dover Publications.

\item Vranas, P. B. [1998]: `Epsilon-ergodicity and the success of equilibrium statistical mechanics', \emph{Philosophy of Science}, \textbf{65}, pp. 688--708.

\item Wallace, D. [2015]: `The Quantitative Content of Statistical Mechanics', \emph{Studies in History and Philosophy of Modern Physics}, \textbf{52}, pp. 285--93.

\item Werndl, C. [2009]: `What Are the New Implications of Chaos for Unpredictability?', \emph{British Journal for the Philosophy of Science}, \textbf{60}, pp. 195--220.

\item Werndl, C. [2013]: `Justifying Typicality Measures of Boltzmannian Statistical Mechanics and Dynamical Systems', \emph{Studies in History and Philosophy of Modern Physics}, \textbf{44}, pp. 470--79.

\item Werndl, C. and Frigg, R. [2015]: `Reconceptionalising Equilibrium in Boltzmannian Statistical Mechanics', \emph{Studies in History and Philosophy of Modern Physics}, \textbf{49}, pp. 19--31.

\item Werndl, C. and Frigg, R. [2017a]: `Mind the Gap: Boltzmannian vs Gibbsian Equilibrium', \emph{Philosophy of Science}, \textbf{84}, pp.  1289--1302.

\item Werndl, C. and Frigg, R. [2017b]: `Boltzmannian Equilibrium in Stochastic Systems', in M. Massimi and J. Romeijn (\emph{eds}),  \emph{Proceedings of the EPSA15 Conference}, Berlin and New York: Springer, pp. 243--54.

\item Werndl, C. and Frigg, R. [forthcoming]: `When does a Boltzmannian equilibrium exist?', forthcoming in D. Bedingham, O. Maroney, and C. Timpson (\emph{eds}), \emph{Quantum Foundations of Statistical Mechanics}, Oxford: Oxford University Press.

\item Werndl, C. and Frigg, R. [unpublished]. `The Limits of Gibbsian Statistical Mechanics', Manuscript. 

\end{list}

\end{document}